\documentclass{article}
\usepackage[utf8]{inputenc}
\usepackage{amsmath}
\usepackage{amsfonts}
\usepackage{amssymb}
\usepackage[english]{babel}
\usepackage{cite}
\usepackage{color}
\usepackage{epsfig}
\usepackage{cite}
\usepackage{graphicx}%
\usepackage{dcolumn}%
\usepackage{bm}%
\begin{document}
	
\begin{center}
{\large \bf{Stability analysis of an ensemble of simple harmonic oscillators}}\\
\vspace{1cm}
R. K. Thakur$^{1}$,
B. N. Tiwari$^{2,3}$,
R. Nigam$^{1}$,
Y. Xu$^{4}$, and
P.K.Thiruvikraman$^{1}$
\vspace{1cm}

$^1$ Department of Physics, Birla Institute of Technology $\&$ Science, Pilani  Hyderabad Campus, Hyderabad- 500078 India.\\
$^2$ INFN-Laboratori Nazionali di Frascati,	Via E. Fermi 40, 00044 Frascati, Rome, Italy.\\
$^3$ University of Information Science and Technology, ``St. Paul the Apostle", Partizanska Str. bb 6000 Ohrid, Republic of Macedonia.\\
$^4$ Department of Applied Mathematics, School of Science, 
Northwestern Polytechnical University, Xi'an, 710072, China

\end{center}
\author{thakurr58 }
\date{November 2018}
\section*{Abstract}
In this paper, we investigate the stability of the configurations of harmonic oscillator potential that are directly proportional to the square of the displacement. We derive expressions for fluctuations in partition function due to variations of the parameters, viz. the mass, temperature and the frequency of oscillators. Here, we introduce the Hessian matrix of the partition function as the model embedding function from the space of parameters to the set of real numbers. In this framework, we classify the regions in the parameter space of the harmonic oscillator fluctuations where they yield a stable statistical configuration. The mechanism of stability follows from the notion of the fluctuation theory. In sections 7 and 8, we provide the nature of local and global correlations and stability regions where the system yields a stable or unstable statistical basis, or it undergoes into geometric phase transitions. Finally, in section $9$, the comparison of results is provided with reference to other existing research. \\

{\it {\bf Keywords:} Stability Analysis, Statistical Configurations, Simple Harmonic Oscillator, Ensemble Fluctuations, Thermodynamics.}\\ 

%
%
\section{Introduction}
Harmonic oscillators are among the important problems that arise in different fields like mechanics and physics. They provide us a useful model for a class of waves and vibrational phenomena that we encounter in classical mechanics and dynamics, statistical mechanics, electrodynamics, solid state physics, atomic physics, nuclear physics and particle physics\cite{Rupp}. In quantum mechanics, it is as an invaluable tool for formulating basic concepts and formalism, for instance, see \cite{Zettili}. There are a number of discussions about the fluctuations of thermodynamic variables of an equilibrium system and the theory of irreversible processes \cite{19a}. 

The motivations and importance of our study are summarized as follows. In this paper, we studied statistical fluctuation to determine stable regions and regions of phase transitions for an ensemble of harmonic oscillators with respect to their parametera $\{T,m,K\}$, where $T$ is temperature $m$ is mass and $K$ is spring constant. There have been various studied involving phase transitions for different systems like thermal fluctuations induced by increasing temperature that can change the state of matter. For example, there is a phase transition when water boils to steam. It also is possible to change the state of matter at absolute zero temperature by means of certain quantum fluctuations  that are demanded by Heisenberg's uncertainty principle. In this case, the quantum phase transitions arise from one state to another that are  provided by adjusting a tuning parameter \cite{Sachdev}. 

In particular, the stability structures of an ensemble of harmonic oscillators are defined by a harmonic potential. Different kinds of stability analysis provide different perspectives. For example, from the perspective of numerical analysis, we consider an intrinsic analysis through the von Neumann stability criteria \cite{7} in the light of information theory. Further, Fourier analysis techniques are exploited to compute the stability of a series of finite differences. Indeed, as the equations of motion, such arrangements are equally applicable to a set of linear partial differential equations. By using a similar analysis, we analyze the local and global stability properties of a given statistical system \cite{Eurocode}. In particular, our motivation is to compute fluctuation invariants such an the correlation length, correlation area and correlation volume of an ensemble of harmonic oscillators. In the sequel, as a particular case, we wish to compute ensemble stability under the fluctuations of the parameters of simple harmonic oscillators. 

The novelty of our work is the consideration of fluctuations in the space of model parameters, viz. the mass, temperature and the frequency of oscillators. Herewith, by introducing the Hessian matrix and  considering the partition function as the model embedding function from the space of parameters to the set of real numbers, we address the issues of stability of an ensemble of harmonic oscillators. The statistical analysis of such an ensemble is an essential point of our work that has not been investigated till date. Indeed, different aspects of the fluctuation theory are performed in the sequel of this paper. In short, we determine stability properties of an ensemble of harmonic oscillators. The explicit analysis is presented in sections $6$ and $7$ respectively. The  discussion and comparison of results are provided in sections $8$ and $9$.

Notice further that the fluctuation theory is brought into the picture in order to prove fundamental properties of irreversible processes and associated reciprocal relations. Namely, as a set of linear equations, this emerges as the symmetry of the fluctuation matrix of the coefficients that relate thermodynamic forces and fluxes. The above connection postulates that the decay of a given system in a chosen non-equilibrium state is produced by spontaneous fluctuations. Empirically, this obeys an average law of the concerned decay from the aforementioned state back to the equilibrium. In case, if it is produced via a constraint, then such fluctuations get suddenly undermined \cite{20a}. 

In the light of quantum fluctuations and information systems, one finds the superconducting devices as promising candidates towards the implementation of qubits (solid-state quantum bits). Indeed, the single-qubit operations \cite{21a, 22a,23a,24a,25} directly couple between two qubits, see \cite{26,27,28,29} for an introduction. Hereby, there exists a physical realization of quantum gates \cite{30}. In this regards, there is a number of complex manipulations in reference to entangled states. On one hand, with reference to the coupling of an arbitrary two-level system to a quantum harmonic oscillator, there are ion/atom-trap experiments, see \cite{31, 32} for the single ion trapping and optical atom trapping respectively. On the other hand, the quantum groups such as $SU(2)_q$ arise analogously to the Schwinger’s  development of the quantum theory of angular momenta. Such a theory is the $q$-analogue of the standard quantum harmonic oscillator \cite{33}.

Fluctuating systems offer a significant new addition to the understanding of the conventional dynamical and mechanical repertoire of limiting equilibrium configurations, e.g., periodic oscillators. Central to their utilities to address the observed random behavior, the concerned stability and the observability of bifurcation sequences are among the important issues. In the light of the present research, we focus on their analysis in  presence of various noises that arise as certain randomness in a system with multiple sources. In particular, with the fact that fluctuations play an important role in dynamical systems, we wish to understand their role viz \`a viz their chaotic nature generated in deterministic nonlinear dynamical systems \cite{9}.

In the light of matrix analysis, the Wigner and Wishart matrices arise as the cornerstone of random matrix theory. Both these matrices find applications in a number of fields of science and engineering \cite{10, 11, 12}. In this regard, concerning quantum fluctuations, the notion of Wigner matrices arises due to Wigner who investigated a certain special class of large dimensional random matrices. This explores the properties of eigenfunctions and eigenvalues for complex quantum mechanical systems. This consideration gives information about the stability of the underlying system\cite{13} under fluctuations of its parameters. In general, such a quantum stability problem is not easy to solve \cite{Rupp}, therefore we purpose to explore the underlying statistical stability in the  space of parameters.
\section{Review of the Model}
Below, we provide a brief review of the thermodynamic fluctuation theory. First of all, in the light of equilibrium thermodynamic fluctuations as described in the Boltzmann-Einstein theory, the probability $\mathcal{P}$ of fluctuations from the equilibrium of a given macroscopic region of volume $V$ is proportional to 
\begin{equation}
\mathcal{P}= \exp{(V \Delta S/k)}, 
\end{equation}
where $\Delta S$ is the variation of the entropy density calculated along with a chosen reversible transformation that causes the fluctuation and $k$ is the standard Boltzmann constant, for example, see Lanford \cite{1} equilibrium states in classical statistical mechanics and fluctuations of the entropy. This theory is based on a rigorous mathematical formulation in the realm of the classical equilibrium statistical mechanics. Notice further that such a model is termed as the theory of large deviations, see \cite{1} for an overview.

Let us recall that a thermodynamic state is represented by a definite value of the pressure $P$, volume $V$ and temperature $T$. For a given system with $(P, V, T)$ as above, its thermodynamic states $f$ are represented by the equation 
\begin{equation} 
f(P,V,T) = 0 
\end{equation}
The fluctuations in the system largely occur if any of these quantities, viz. $(P, V, T)$ varies, that is, their values do not remain fixed. Such fluctuations depend on the type of the system and its concerned thermodynamic states. Basically, fluctuations are understood as a deviation of the system in equilibrium from its average state \cite{8}. In particular, the fluctuation theory \cite{8} arises in various branches of physics including thermodynamics, statistical mechanics, viz. energy fluctuations, entropy fluctuations, particle number fluctuations and others. We have considered the potential fields in which the system fluctuates with mass, spring constant or frequency and temperature. This is realized by defining the partition function for an isolated system. Then, the partition function is extremized to see the nature of fluctuations with respect to parameters under their variations.

Despite the fact that the thermodynamics fluctuation theory originates from statistical mechanics, it may be considered on another basis that is completely thermodynamical, where there is no essential need of any microscopic foundation to understand fluctuations \cite{2}. The importance of fluctuations arises further in the realm of localization theories and quantum transport phenomena, as well. By the thermodynamic limit, we mean $N \rightarrow \infty $, where $N$ is the number of particles. In this case, it follows that the underlying fluctuations in the ensemble vanish. Nonetheless, earlier studies include fluctuations about an equilibrium of one-dimensional problems, for example, see \cite{3} towards the diffusive behavior of a metallic system and its quantum aspects. In general, for stationary non-equilibrium states, dynamical fluctuation theory gives explicit tests of stochastic models for an ensemble of interacting particles. In this framework, the time-reversed dynamics plays a crucial role, see \cite{4} for an overview. Hereby, in terms of the microscopic degree of freedom of a chosen configuration, let us recall \cite{1} that the famous formula of Boltzmann 
\begin{equation} 
S= k_B \ln \omega, 
\end{equation}  
defines the macroscopic entropy through the second law of thermodynamics. Hereby, Einstein gave an ingenious idea to invert the above relation in order to calculate microscopic probabilities from thermodynamics of a given system \cite{5}.
\section{Thermodynamics via Partition Function}
With the motivations that the partition function gives the essential relation between the coordinates of microscopic systems and the thermodynamic properties. Physically, a partition function explains the statistical properties of a system which are in thermodynamic equilibrium. In general, partition functions are functions of the thermodynamic state variables such as the temperature, pressure and volume. Many of the thermodynamic variables of the system can be stated in terms of the partition function or its derivatives, for example the total energy, free energy, entropy, and pressure of the system \cite{5}. The partition function of a definite physical system has no dimension.

As mentioned above, once the partition function is evaluated, we can calculate all the physical quantities such as internal energy $U$, entropy $S$, specific heats at constant pressure and constant volume, viz. $C_P$ and $C_V$ respectively, see \cite{courses} for an overview. Geometrically, we find that the partition function can be realized as the volume occupied by the system in its phase space. Basically, it tells us how many microstates of a system are accessible in a given ensemble. This can be seen directly by starting from the microcanonical ensemble, see \cite{stackexchange} for the physical meaning of a partition function in statistical physics.

Although the information theory is largely regarded as a branch of communication theory, it spans a number of disciplines including physics, statistics, computer science, probability, economics, and others. The fundamental questions that are treated by information theory include (i) how can the information be measured? (ii) How can the information be transmitted? and (iii) How can the information be stored? From the communication theory viewpoints,  we may assume that the information is carried either by certain signals, or symbols. Largely, this is an open problem. In the case of the Shannon's sampling theory, our consideration suggests that if a channel is band-limited, we can consider samples of the signal without any information loss. 

On the other hand, Shannon’s consideration anticipates that information is a measure of the randomness. Subsequently, information sources are modeled by various random processes, whose statistical invariants largely depend on the nature of the chosen information sources \cite{7}. In this regard, the associated fluctuation theorems \cite{14} offer an analytic expression of the probability. In a nonequilibrium configuration of finite size that is observed for a finite duration of time, the dissipative flux flows in the reverse direction to that the one required by the second law of thermodynamics. In the sequel, we find that the same has been confirmed numerically in various cases, see \cite{14,15,16,17,18,19}.

Recently, there has been a number of studies towards the stability criteria of classical and quantum systems. In this framework, the fluctuation theory provides a constructive criterion for setting up probability distributions with partial knowledge. This leads to an intrinsic statistical inference model termed as the maximum entropy estimation problem. Practically, this results into the least biased estimated system on a given information basis. In other words, this offers the maximally noncommittal entropy with respect to some missing information. If one considers that statistical mechanics arising from an inference model rather than a physical theory, it follows that the usual computational rules, starting from the determination of the partition function, are immediate consequences of the entropy maximization principle \cite{6}.

In the case of physical theories, a partition function describes the statistical properties of a system in thermodynamic equilibrium \cite{8}. Partition functions can be viewed as functions of the thermodynamic state variables, such as the temperature $T$, pressure $P$ and volume $V$. Most of the average thermodynamic variables of the system, viz. the total energy $E$, entropy $S$, free energy $F$, and pressure $P$, can be derived directly from the partition function or its certain derivatives \cite{patharia}.
The partition function is equally used in statistical mechanics, both in its classical and quantum counterparts. On the other hand, solid state, low temperature systems, low pressure processes, and most of the chemical configurations depend on the equipartition function. It is known that the partition function also yields a direct proof for the equipartition theorem. Recently, it has been found that the equipartition theorem finds a broad range of uses in chemistry, physics, chemical engineering and other applied sciences \cite{9}.
\section{Methodology}
In this section, the definition and solution based on the method of extrema are listed. Subsequently, we consider correlations in an ensemble of harmonic oscillators by introducing randomness in its parameters. It is worth mentioning that fluctuations are among the main causes because of which we have variations in the position of a particle. Hereby, we consider parameter space fluctuations. This relies on a general technique for adequately finding the maxima and minima of the underlying partition function as the model embedding function, see \cite{bt} towards the state-space correlations and stabilities of black hole configurations.

First of all, recall that the value of the function at a maximum point is called the maximum value of the function. Correspondingly, the value of the function at a minimum point is called the minimum value of the function. In the mathematical analysis, the maxima and minima of a function, collectively known as extrema, are the largest and the smallest value of the function. That is, within a given range, the local or relative extrema or on the entire domain of the function are termed as the global or absolute extrema. With the help of this, we can locate points on a graph where the associated gradient vanishes. 

In this work, we show that such points are associated with the largest or smallest values of the partition function. Such a consideration holds at least in their immediate vicinity in the space of parameters. In many applications, scientists, engineers, or economists are interested in finding such points for various reasons such as the maximization of the power, or profit, or minimization of losses or cost functions. Physically, at such points at which the tangent to the graph of the objective function is horizontal are called its stationary points. Therefore, we can locate stationary points by looking at the points at which its slope vanishes. In this concern, we examine the algebraic and geometric properties concerning the stability of an ensemble of harmonic oscillators.
\section{Calculation of Partition Function}
In this section, we provide an explicit calculation of the above mentioned problem. Let
$ N$ be number of classical particles that are confined in a one dimensional harmonic oscillator with the total energy
\begin{equation} 
 E = \frac{p_x^2}{2m}+\frac{1}{2} mw^2 x^2,
\end{equation} 
where  $x$ is the position, m $p_x$ is the conjugate momentum, $m$ is the mass of the oscillator and $w$ is the corresponding angular frequency of oscillations. Here, the first term denotes the kinetic energy and the second is the potential energy of the oscillator. 
In this case, the one dimensional harmonic oscillator partition function is defined as
\begin{equation} Z_1 = \frac{1}{h} \int^{+\infty}_{-\infty} \int^{+\infty}_{-\infty} \exp{(-\frac{p_x^2+ m^2w^2 x^2}{2mk_BT})} dx dp_x
\end{equation} 
Hereby, it follows that the respective partition function is given by
\begin{equation} \label{onepart}
Z_1= \frac{2 \pi k_B T}{hw}
\end{equation} 
Physically, $Z_1$ is the partition function of the single classical particle.
Similarly, extending the case of the above one dimensional harmonic oscillator to arbitrary $N$ dimensions, with the spring constant $K= mw^2$, 
we find that the above partition function reads as
\begin{equation} \label{npart}
Z_N = \frac{1}{N!h^N}(2\pi k_B T)^N(\frac{m}{K})^\frac{N}{2}
\end{equation} 
Notice that Eqn.(\ref{npart}) is obtained from Eqn.(\ref{onepart}) by considering an $N$ particle harmonic oscillator configuration whose partition function is defined as
$Z_N= Z_1^N/N!$. In order to study the parametric stability of harmonic oscillator configurations, we randomize the oscillator parameters $\{T, m, K\}$ with a randomized partition function. The concerned randomized partition function $Q_N$ as thermodynamically equivalent to $Z_N$ when its parameters $\{T, m, K\}$ fluctuate. In this case, under variations of $\{T, m, K\}$, we define the fluctuation matrix as
\begin{equation}    \label{fmat}
M=
  \left({\begin{array}{ccc}
   \frac{\partial^2Q_N}{\partial T^2} & \frac{\partial^2Q_N}{\partial T \partial m} & \frac{\partial^2Q_N}{\partial T \partial K}  \\
   \frac{\partial^2Q_N}{\partial T \partial m} & \frac{\partial^2Q_N}{\partial m^2} & \frac{\partial^2Q_N}{\partial m \partial K}\\
   \frac{\partial^2Q_N}{\partial T \partial K} & \frac{\partial^2Q_N}{\partial m \partial K} & \frac{\partial^2Q_N}{\partial K^2}
  \end{array} } \right),
\end{equation} 
where $\frac{\partial^2Q_N}{\partial x_i \partial x_j}$ are the second order partial derivatives of the limiting randomized partition function $Q_N$ with respect to the model parameters $x_i$ and $x_j$.
\section{Local Fluctuations and Correlation}
In this section, we compute the components of the fluctuation matrix $M$ as in Eqn.(\ref{fmat}). Hereby, the associated principal minors are calculated in order to judge the stability of the ensemble under fluctuations of its parameters. Following the above definition as in  Eqn.(\ref{npart}), we have the following flow components\\  
\begin{eqnarray} 
\frac{\partial Q_N}{\partial T} &=& \frac{(2 \pi K_B)^N}{N!h^N} (\frac{m}{K})^\frac{N}{2} N T^{N-1}, \nonumber \newline\\
\frac{\partial Q_N}{\partial m} &=& \frac{(2 \pi K_B T)^N}{N! h^N} \frac{N m^\frac{N-2}{2}}{2K^\frac{N}{2}}, \nonumber \newline\\
\frac{\partial Q_N}{\partial K} &=& \frac{(2 \pi K_B T)^N}{N!h^N} (m)^\frac{N}{2} (\frac{-N}{2}) \frac{1}{K^\frac{N+2}{2}}
\end{eqnarray}

The needed statistical auto and cross corrections are
\begin{eqnarray} 
\frac{\partial^2Q_N}{\partial T^2} &=& \frac{(2 \pi K_B)^N}{N!h^N} (\frac{m}{K})^\frac{N}{2} N (N-1) T^{N-2}, \nonumber \newline\\
\frac{\partial^2Q_N}{\partial m^2} &=& \frac{1}{2N!h^N} (2 \pi K_BT)^N \frac{N}{(K)^\frac{N}{2}} (\frac{N-2}{2}) (m^\frac{N-4}{2}), \nonumber \newline\\
\frac{\partial^2Q_N}{\partial K^2} &=& \frac{(2 \pi K_B  T)^N}{N!h^N} (m)^\frac{N}{2} (\frac{1}{(K)^\frac{N+4}{2}}) (\frac{-N}{2}) (\frac{N+2}{2}), \nonumber \newline\\
\frac{\partial^2Q_N}{\partial T \partial K} &=& \frac{-N^2}{2N!h^N} N^2 (2 \pi K_B T)^N T^N-1 \frac{(m)^\frac{N}{2}}{(K)^\frac{N+2}{2}}, \nonumber \newline\\
\frac{\partial^2Q_N}{\partial m \partial K} &=& \frac{-N^2}{4N!h^N} (2 \pi K_B T)^N \frac{(m)^\frac{N-2}{2}}{(K)^\frac{N+2}{2}}, \nonumber \newline\\
\frac{\partial^2Q_N}{\partial T \partial m} &=& \frac{N^2}{2N!h^N} (2 \pi K_B)^N T^N-1 \frac{m^\frac{N-2}{2}}{K^\frac{N}{2}}
\end{eqnarray}

To simplify the calculations, let us write the above fluctuation matrix as in Eqn.(\ref{fmat}) as shown below 
\begin{equation}
   M=   \left( {\begin{array}{ccc}
   A & B & C  \\
   B & D & E\\
   C & E & F
  \end{array} } \right)
\end{equation}
In order to classify the system stability under parametric fluctuations, we have employed the components of $M$ according to the results as below. In the sequel, we examine the parameter space stability of a sample of harmonic systems under their fluctuations.

\section{System Stability}
In this section, the associated principal minors are calculated in order to judge the stability of the considered ensemble of harmonic oscillators under fluctuations of parameters. Following the above definition, we compute the principal minors of $M$ as follows. Given the $3 \times 3$ symmetric matrix $M$ as above, it follows that $M$ has the following principal minors 
\begin{eqnarray} 
m_1&:=& A,  \nonumber \newline\\
m_2&:=& AD-B^2  \nonumber \newline\\
m_3&:=& A(DF-E^2)-B(BF-CE)+C(BE-CD)
\end{eqnarray} 
With the above expressions of second derivatives of $Z_N$, the principal minors are given as the below expressions
\begin{eqnarray} 
m_1&:=& \frac{(2 \pi K_B)^N}{N!h^N} (\frac{m}{K})^\frac{N}{2} N(N-1) T^{N-2}, \nonumber \newline\\ 
m_2&:=& \frac{1}{2}(\frac{N^2}{N!h^N})^2 (2 \pi K_B)^2N  T^2N-2   m^{N-2} (\frac{1}{K})^N  (\frac{N^2-3N+1}{2}), \nonumber \newline\\
m_3&:=& (\frac{1}{N!h^N})^3 (2 \pi K_B)^{3N} (T)^{3N-2} \frac{m^{\frac{3N-4}{2}}}{K^{\frac{4+3N}{2}}} (\frac{2-N^2+N^4+N^5-N^6}{8})
\end{eqnarray} 
\section{Discussion of the Results}
In order that the ensemble remains stable, we require the positivity of all the principal minors $\{m_i \ | \ i = 1,2,3\}$. This is possible when the principal minor $m_3$ is greater than zero, that is, we have $n_3(N):=2-N^2+N^4+N^5-N^6 >0$. This is an equation of degree six in $N^6$. 

\begin{center}
\begin{figure}
\hspace*{2.0cm} \vspace*{-0.5cm}
\includegraphics[width=8cm,angle=0]{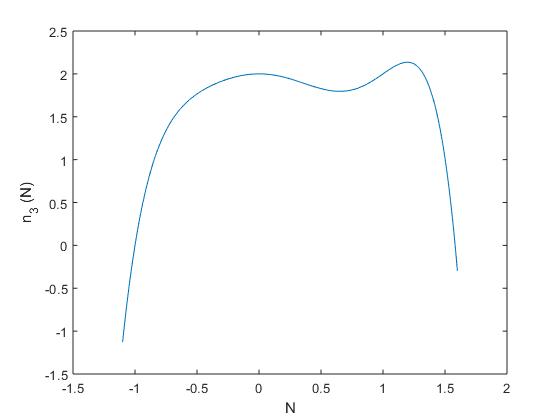}
\caption{The $n_3(N)$ plotted as a function of the number of particles $N$ in the ensemble plotted on $Y$-axis and $N$ on $X$-axis describing the nature of roots by considering variations of the number of particles in the ensemble.} \label{fig}
\vspace*{-0.5cm}
\end{figure}
\end{center}

Numerically, we observe that there is a local minimum near $n=0.5$ that indicates a complex root of $n_3(N)=0$, see figure (\ref{fig}). After solving the equation $n_3(N)=0$ explicitly by Newton-Raphson method \cite{NR}, with an iteration of $i$ used as $0.25+i$ as an initial guess, we observe that there are two real roots $-1$ and $1.5937$ of $n_3=0$. Further, there are two pairs of complex roots of $n_3=0$ that read as $0.803625173811080 \pm 0.693303740971437 i \in \mathbb{C}$ and $-0.594425456895094 \pm 0.877049136315012i$. Notice that a complex roots of $n_3(N)=0$ signifies damping in the system. This is in parallel to the fact that a simple harmonic oscillator with complex frequency of oscillations becomes a damped harmonic oscillator.

In short, there are total of two real roots and four complex roots of $n_3=0$. We may equally choose different starting values to converge to the above root in a given numerical scheme.
Similarly, the positivity of the principal minor $m_2$ requires that the dimension $N$ satisfies a quadratic inequality $N^2-3N+1>0$, that is, $N$ is out of the interval $((3-\sqrt{5})/2, (3+\sqrt{5})/2)\simeq (0.2, 2.8)$. Thus, under the quadratic condition that $m_2>0$, a harmonic oscillator ensemble is unstable when its natural dimension takes value in $\{1, 2, 3\}$ over $\mathbb{N}$. Finally, we see that the positivity of the principal minor $m_1$ requires that $N \notin (0, 1)$.
\section{Comparison with Other Existing Research}

A comparison with other existing results is as follows. A simple harmonic oscillator is stable when the damping constant is zero. In other words, if the frequency of oscillations is an imaginary valued quantity, it will be damping in the system \cite{patharia, 8}. Simple harmonic oscillator has no decay, that is, it corresponds to a stable system having oscillatory motion. Thus, the instability arises in the system when a decay factor or damping parameter is added to the simple harmonic oscillator. 

A similar analysis is followed for driven harmonic oscillators in an externally applied force. In this case, the amplitude and phase of oscillations govern the behaviour of motion in order for matching the initial conditions. Here, we focus on the stability of an ensemble of harmonic oscillators by invoking the Hessian matrix approach to examine the stability and instability domains in the space of its parameters. The oscillators thus considered correspond to parametric harmonic oscillators that are certain driven simple harmonic oscillators, see \cite{pho} in the light of optical parametric oscillators. Notice that parametric excitations are different from forcing because the action turns out to be a time varying modification on parameters of the system. 

Our analysis follows standard statistical mechanical framework that is based on the dimension of phase-space. In this regard, in section $8$, we have shown in which dimensions of the phase space, the underlying ensemble of particles is stable. Hereby, as discussed in section $8$, the signature of principal minor $\{ p_1, p_2, p_3 \}$ shows whether there is damping or not in the considered ensemble of oscillators. Similar kind of works have been done for the black holes with one or multiple centers \cite{bt}, and cosmic ray particles \cite{Ourgluonpaper}.

\section{Conclusion}
In this paper, we have investigated the stability analysis of configurations in the harmonic oscillator limit. Namely,  we introduce Hessian matrix by considering the partition function as the model embedding function from the space of parameters to the set of real numbers. Hereby, we have examined the stability of configuration of the harmonic oscillator potential. In doing so, the thermodynamic fluctuation theory is summarized. Following the same, we have focused on the fluctuations of the partition function in order to analyze the limiting thermodynamic properties.  In this setup, we have determined the domains of stability of an ensemble of harmonic oscillators. 

Subsequently, by using numerical techniques, we notice that a local minimum near $n=0.5$ indicates a pair of complex factors of the third minor. We have solved the equation $n_3(N)=0$ explicitly by using the Newton Raphson method. As a result, we observe that there are four complex roots and two real roots for the equation $n_3(N)=0$.  This provides a detailed account of parametric fluctuations. In addition, we have performed different starting values to converge to the above sets of real or complex roots $n_3(N)$.

This paper presents the concerned calculation in the next three parts, local stability, local correlations and system stability. Our results provide the criteria for stability of an ensemble of harmonic oscillators. In the above framework, the perspective directions include stochastic optimization of statistical ensembles and their stabilities under variations of the model parameters.


\begin{thebibliography}{1}

\bibitem{Rupp} Ruppeiner, G., ``Riemannian geometry in thermodynamic fluctuation theory." Rev. Mod. Phys, 67, no. 3 (1995) 605, [Erratum, 313-313, 68].

\bibitem{Zettili} Zettili, Nouredine. "Quantum mechanics: concepts and applications." (2003): 93-93.

\bibitem{19a} Onsager, Lars. "Reciprocal relations in irreversible processes. I." Physical review 37.4 (1931): 405.

\bibitem{Sachdev} Sachdev, S., Quantum phase transitions. Handbook of Magnetism and Advanced Magnetic Materials, 2007.

\bibitem{7} Witten, Edward. "A mini-introduction to information theory." arXiv preprint arXiv:1805.11965 (2018).


\bibitem{Eurocode} Code, Price. "Eurocode 8: Design of structures for earthquake resistance-part 1: general rules, seismic actions and rules for buildings." Brussels: European Committee for Standardization (2005).

\bibitem{20a} Onsager, Lars, and S. Machlup. "Fluctuations and irreversible processes." Physical Review 91.6 (1953): 1505.

\bibitem{21a} Nakamura, Yu, Yu A. Pashkin, and J. S. Tsai. "Coherent control of macroscopic quantum states in a single-Cooper-pair box." nature 398.6730 (1999): 786.

\bibitem{22a} Vion, Denis, et al. "Manipulating the quantum state of an electrical circuit." Science 296.5569 (2002): 886-889.

\bibitem{23a} Yu, Yang, et al. "Coherent temporal oscillations of macroscopic quantum states in a Josephson junction." Science 296.5569 (2002): 889-892.

\bibitem{24a} Martinis, John M., et al. "Rabi oscillations in a large Josephson-junction qubit." Physical review letters 89.11 (2002): 117901.

\bibitem{25} Chiorescu, I., et al. "Coherent quantum dynamics of a superconducting flux qubit." Science 299.5614 (2003): 1869-1871.

\bibitem{26} Pashkin, Yu A., et al. "Quantum oscillations in two coupled charge qubits." Nature 421.6925 (2003): 823.

\bibitem{27} Berkley, A. J., et al. "Entangled macroscopic quantum states in two superconducting qubits." Science 300.5625 (2003): 1548-1550.

\bibitem{28} Majer, J. B., et al. "Spectroscopy on two coupled flux qubits." arXiv preprint cond-mat/0308192 (2003).

\bibitem{29} Chiorescu, I., et al. "Coherent dynamics of a flux qubit coupled to a harmonic oscillator." Nature 431.7005 (2004): 159.

\bibitem{30} Yamamoto, Tsuyoshi, et al. "Demonstration of conditional gate operation using superconducting charge qubits." Nature 425.6961 (2003): 941.

\bibitem{31} Leibfried, Dietrich, et al. "Quantum dynamics of single trapped ions." Reviews of Modern Physics 75.1 (2003): 281.

\bibitem{32} Mandel, Olaf, et al. "Controlled collisions for multi-particle entanglement of optically trapped atoms." Nature 425.6961 (2003): 937.

\bibitem{33} Macfarlane, A. J. "On q-analogues of the quantum harmonic oscillator and the quantum group SU (2) q." Journal of Physics A: Mathematical and general 22.21 (1989): 4581.

\bibitem{1}	Lanford, Oscar E. "Entropy and equilibrium states in classical statistical mechanics." Statistical mechanics and mathematical problems. Springer, Berlin, Heidelberg, 1973. 1-113.
 
\bibitem{2} Ruppeiner, George. "Riemannian geometry in thermodynamic fluctuation theory." Reviews of Modern Physics 67.3 (1995): 605.

\bibitem{3} Muttalib, K. A., J-L. Pichard, and A. Douglas Stone. "Random-matrix theory and universal statistics for disordered quantum conductors." Physical review letters 59.21 (1987): 2475.

\bibitem{4} Bertini, Lorenzo, et al. "Macroscopic fluctuation theory for stationary non-equilibrium states." Journal of Statistical Physics 107.3-4 (2002): 635-675.

\bibitem{5} Eyink, Gregory L. "Dissipation and large thermodynamic fluctuations." Journal of statistical physics 61.3-4 (1990): 533-572.

\bibitem{6} Jaynes, Edwin T. "Information theory and statistical mechanics. II." Physical review 108.2 (1957): 171.

\bibitem{patharia} Patharia, R. K. "Statistical Mechanics, 2nd Edition." (1996).

\bibitem{8} Huang, Kerson. Introduction to statistical physics. Chapman and Hall/CRC, 2009.

\bibitem{9}	Crutchfield, James Patrick, J. Doyne Farmer, and Bernardo A. Huberman. "Fluctuations and simple chaotic dynamics." Physics Reports 92.2 (1982): 45-82.

\bibitem{courses} Callen, Herbert B. "Thermodynamics and an Introduction to Thermostatistics." (1998): 164-167.

\bibitem{stackexchange} Kadanoff, Leo P. Quantum statistical mechanics. CRC Press, 2018.

\bibitem{10} Kumar, Santosh. "Random matrix ensembles involving Gaussian Wigner and Wishart matrices, and biorthogonal structure." Physical Review E 92.3 (2015): 032903.

\bibitem{11} Mehta, Madan Lal. Random matrices. Vol. 142. Elsevier, 2004.

\bibitem{12} Akemann, Gernot, Jinho Baik, and Philippe Di Francesco. The Oxford handbook of random matrix theory. Oxford University Press, 2011.

\bibitem{13} Wigner, Eugene P. "On the distribution of the roots of certain symmetric matrices." Ann. Math 67.2 (1958): 325-327.

\bibitem{14} Ayton, Gary, Denis J. Evans, and Debra J. Searles. "A local fluctuation theorem." The Journal of chemical physics 115.5 (2001): 2033-2037.

\bibitem{15} Evans, Denis J., and Debra J. Searles. "Steady states, invariant measures, and response theory." Physical Review E 52.6 (1995): 5839.

\bibitem{16} Lepri, S., L. Rondoni, and G. Benettin. "The Gallavotti–Cohen fluctuation theorem for a nonchaotic model." Journal of Statistical Physics 99.3-4 (2000): 857-872.

\bibitem{17} Bonetto, F., G. Gallavotti, and P. L. Garrido. "Chaotic principle: an experimental test." Physica D: Nonlinear Phenomena 105.4 (1997): 226-252.

\bibitem{18}  Bonetto, F., N. I. Chernov, and J. L. Lebowitz. "(Global and local) fluctuations of phase space contraction in deterministic stationary nonequilibrium." Chaos: An Interdisciplinary Journal of Nonlinear Science 8.4 (1998): 823-833.

\bibitem{19} Evans, Denis, Gary Ayton, and Debra J. Searles. "A Local Fluctuation Theorem." (1999).

\bibitem{20} Onsager, Lars. "Reciprocal relations in irreversible processes. I." Physical review 37.4 (1931): 405 and Onsager, Lars. "Reciprocal relations in irreversible processes. II." Physical review 38.12 (1931): 2265.

\bibitem{21} Onsager, Lars, and S. Machlup. "Fluctuations and irreversible processes." Physical Review 91.6 (1953): 1505.

\bibitem{22} Nakamura, Yu, Yu A. Pashkin, and J. S. Tsai. "Coherent control of macroscopic quantum states in a single-Cooper-pair box." nature 398.6730 (1999): 786.

\bibitem{23} Vion, Denis, et al. "Manipulating the quantum state of an electrical circuit." Science 296.5569 (2002): 886-889.

\bibitem{24} Yu, Yang, et al. "Coherent temporal oscillations of macroscopic quantum states in a Josephson junction." Science 296.5569 (2002): 889-892.

\bibitem{25} Martinis, John M., et al. "Rabi oscillations in a large Josephson-junction qubit." Physical review letters 89.11 (2002): 117901.

\bibitem{26} Chiorescu, I., et al. "Coherent quantum dynamics of a superconducting flux qubit." Science 299.5614 (2003): 1869-1871.

\bibitem{27} Pashkin, Yu A., et al. "Quantum oscillations in two coupled charge qubits." Nature 421.6925 (2003): 823.

\bibitem{28} Berkley, A. J., et al. "Entangled macroscopic quantum states in two superconducting qubits." Science 300.5625 (2003): 1548-1550.

\bibitem{29} Majer, J. B., et al. "Spectroscopy on two coupled flux qubits." arXiv preprint cond-mat/0308192 (2003).

\bibitem{30} Chiorescu, I., et al. "Coherent dynamics of a flux qubit coupled to a harmonic oscillator." Nature 431.7005 (2004): 159.

\bibitem{31} Yamamoto, Tsuyoshi, et al. "Demonstration of conditional gate operation using superconducting charge qubits." Nature 425.6961 (2003): 941.

\bibitem{32} Leibfried, Dietrich, et al. "Quantum dynamics of single trapped ions." Reviews of Modern Physics 75.1 (2003): 281. 

\bibitem{33} Mandel, Olaf, et al. "Controlled collisions for multi-particle entanglement of optically trapped atoms." Nature 425.6961 (2003): 937.

\bibitem{bt} State-space Correlations and Stabilities, Bellucci, S., Tiwari, B. N. (2010). Phys. Rev. D 82  084008. e-print: arXiv:0910.5309v1 [hep-th], doi.org/10.1103/PhysRevD.82.084008

\bibitem{34} Macfarlane, A. J. "On q-analogues of the quantum harmonic oscillator and the quantum group SU (2) q." Journal of Physics A: Mathematical and general 22.21 (1989): 4581.

\bibitem{NR} Ryaben'kii, Victor S., and Semyon V. Tsynkov. A theoretical introduction to numerical analysis. Chapman and Hall/CRC, 2006.

\bibitem{pho} Byer, R. L. (1975). Optical parametric oscillators. Quantum Electronics: A Treatise, 1(part B).

\bibitem{Ourgluonpaper} Thakur, R.K., Tiwari, B.N. and Nigam, R., 2019. Stability of gluonic systems with multiple soft interactions. Journal of Astrophysics and Astronomy, 40(4), p.34.


\end{thebibliography}
\end{document}